\documentclass[preprint,12pt]{aastex}

\shorttitle{Molecular Bullets in the HH 80--81 Region}
\shortauthors{Qiu \& Zhang}
\begin{document}

\title{Discovery of Extremely High Velocity ``Molecular Bullets'' in
the HH 80--81 High-Mass Star-Forming Region}

\author{Keping Qiu}
\affil{Department of Astronomy, Nanjing University, Nanjing
210093, China} \affil{Harvard-Smithsonian Center for Astrophysics,
60 Garden Street, Cambridge, MA, USA} \email{kqiu@cfa.harvard.edu}

\author{Qizhou Zhang}
\affil{Harvard-Smithsonian Center for Astrophysics, 60 Garden
Street, Cambridge, MA, USA} \email{qzhang@cfa.harvard.edu}

\begin{abstract}
We present Submillimter Array 1.3 mm waveband continuum and
molecular line observations of the HH 80--81 high-mass
star-forming region. The dust continuum emission reveals two
dominant peaks MM1 and MM2, and line emission from high-density
tracers suggests the presence of another core MC. Molecular line
emission from MM1, which harbors the exciting source of the HH
80--81 radio jet, yields a hot molecular core at a gas temperature
of 110 K. The two younger cores MM2 and MC both appear to power
collimated CO outflows. In particular, the outflow arising from
MM2 exhibits a jet-like morphology and a broad velocity range of
190 km\,s$^{-1}$. The outflow contains compact and fast moving
molecular clumps, known as ``molecular bullets'' first discovered
in low-mass class 0 protostellar outflows. These ``bullets''
cannot be locally entrained or swept up from the ambient gas, but
are more likely ejected from the close vicinity of the central
protostar. The discovery of this remarkable outflow manifests an
episodic, disk-mediated accretion for massive star formation.
\end{abstract}

\keywords{ISM: individual (HH 80--81) --- ISM: jets and outflows
--- circumstellar matter --- stars: formation --- stars:
early-type --- masers}

\section{Introduction}
Over the last decade, there have been a growing number of
high-angular-resolution studies using observations from submm/mm
and cm interferometers, aimed at understanding morphological and
kinematical features associated with the massive star formation
process, such as molecular outflows, possible rotating disks or
envelopes, and inward accretion flows \citep[e.g.,][]{Zhang98,
Shepherd98, Beuther02, Cesaroni05, Keto06, Qiu09}. More recently,
studies based on observations from the Submillimeter Array (SMA),
a combined power of high-angular-resolution and broad frequency
coverage, have begun to shed light on the physical and chemical
diversities in high-mass star-forming cores at different
evolutionary stages (e.g., Rathborne et al. 2008; Beuther et al.
2009; Zhang et al. 2009). These efforts provided a wealth of
information that greatly advanced our knowledge of massive star
formation.

In this paper we present a high-angular-resolution study of the
high-mass star-forming region HH 80--81, which has a bolometric
luminosity of $20,000$ $L_{\odot}$ at an adopted distance of 1.7
kpc \citep{Rodriguez80}. This region is distinguished by a 5.3 pc
radio jet that has an opening angle of $\sim$1$^{\circ}$, known as
the best collimated radio jet driven from a high-mass protostar
\citep{Marti93, Marti95, Marti98}. Recent {\it Spitzer} 8 $\mu$m
imaging reveals a bi-conical outflow cavity surrounding the radio
jet \citep{Qiu08}. The jet powering source has been detected from
mid-infrared to cm wavelengths \citep{Aspin94, Stecklum97,
Marti99, Gomez03, Qiu08}, yet some basic properties of the source,
such as physical and chemical conditions of the circumstellar
matter, remain unclear. Here we present a detailed study of the
dust continuum and molecular line emission of the radio jet
powering source, and the discovery of an exceptional CO outflow
containing fast moving molecular bullets.

\section{Observations and Data Reduction}
The SMA\footnote{The Submillimeter Array is a joint project
between the Smithsonian Astrophysical Observatory and the Academia
Sinica Institute of Astronomy and Astrophysics and is funded by
the Smithsonian Institution and the Academia Sinica.} observations
toward the phase center (R.A., decl.)$_{\mathrm
J2000}=(18^{\mathrm h}19^{\mathrm m}12.\!^{\mathrm s}09,
-20^{\circ}\\47^{'}30.\!^{''}9)$ were undertaken with eight
antennas on 2008 June 30 in the compact configuration and on 2008
September 9 in the sub-compact configuration, both under good
weather conditions. Particularly the sub-compact data were taken
with $\tau_{225 \mathrm{GHz}}$$\sim$0.02--0.03. The 4 GHz
correlator was configured to cover rest frequencies of 219.2 to
221.2 GHz in the lower sideband and 229.2 to 231.2 GHz in the
upper sideband. In the compact configuration, we observed Uranus,
3c454.3, and 3c273 for bandpass calibration, Callisto for flux
calibration, and J1911-201 to monitor the time dependent gain. In
the sub-compact configuration, observations of 3c454.3, Uranus,
and J1733-130 and J1911-201 were used for the bandpass, flux, and
gain calibrations, respectively. The data were calibrated using
the IDL MIR package and then exported to MIRIAD for further
processing. Self-calibration was performed on continuum and the
solutions were applied to lines as well. The two data sets were
jointly imaged with the ROBUST parameter set to 0.5, resulting in
a synthesized beam of $\sim$$3.\!''5$${\times}$$2.\!''8$ at FWHM.

\section{Results}
\subsection{Dense Molecular Cores}
\begin{figure}
\epsscale{1.} \plotone{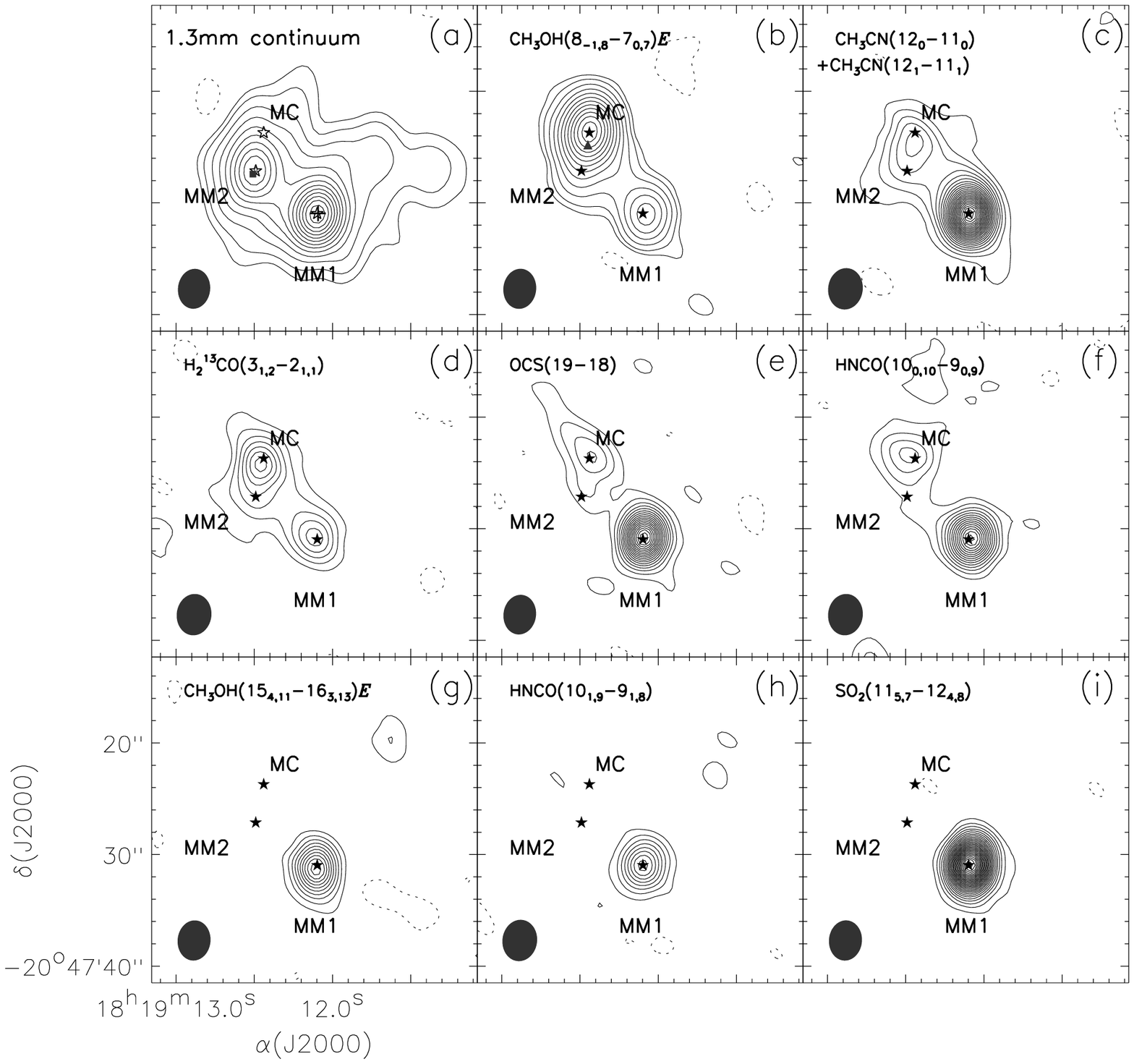} \caption{(a) The 1.3 mm continuum
emission, contouring at ($-1$, 1, 2, 4, 7, 11, 16, 22, 29, 37, 46,
56, 67, 79)${\times}$0.006 Jy\,beam$^{-1}$; a filled square
denotes the brightest H$_2$O maser spot from Marti et al. (1999);
a plus sign marks the position of the central radio source from
Marti et al. (1999). (b)--(i) The molecular line emission,
integrated from 9.6 to 25.2 km\,s$^{-1}$ and contouring at ($-1$,
1, 2, 4, 7, 11, 16, 22, 29, 37, 46, 56, 67)${\times}$0.26
Jy\,beam$^{-1}$\,km\,s$^{-1}$ for CH$_3$OH
(8$_{-1,8}$--$7_{0,7})E$, and integrated from 10.8 to 16.8
km\,s$^{-1}$ and contouring at ($-1$, 1, 2, 3, ...)${\times}$0.12
Jy\,beam$^{-1}$\,km\,s$^{-1}$ for the other lines; a filled
triangle in panel (b) denotes a 44 GHz methanol maser spot from
\citet{Kurtz04}; the line transition is labelled in the upper left
of each panel. Hereafter the three stars denote the peak positions
of the three cores and a filled ellipse in the lower left
delineates the FWHM of the corresponding synthesized beam.
\label{fig1}}
\end{figure}

Figure \ref{fig1} shows the 1.3 mm continuum map and the
integrated intensity maps of lines detected from a variety of
molecular species. The mm continuum reveals two dominant peaks MM1
and MM2; MM1 coincides well with the radio source that powers the
HH 80--81 radio jet and MM2 is spatially associated with H$_2$O
masers and a very weak radio source \citep{Gomez95, Marti99}. In
addition, emission lines from high-density tracers, such as
CH$_3$OH, CH$_3$CN, H$_2\!^{13}$CO, OCS, and HNCO, clearly reveals
another dense core MC (Figures \ref{fig1}b--\ref{fig1}f), although
the mm continuum map only shows a weak enhancement toward this
core. In Figure \ref{fig1}b the CH$_3$OH (8$_{-1,8}$--$7_{0,7})E$
emission toward MC is eminently bright with a peak flux of 6.4
Jy\,beam$^{-1}$. This narrow spectral component at a LSR velocity
of $13.2$ km\,s$^{-1}$ appears to be a maser emission.
\citet{Kurtz04} detected a 44 GHz methanol maser in the vicinity
of MC, which is consistent with our 229.8 GHz detection as both
masers are class I transitions \citep{Slysh02}. Apart from the
maser emission, most thermal line emission in Figure \ref{fig1}
appears brighter toward MM1. In particular lines of CH$_3$OH
(15$_{4,11}$--$16_{3,13})E$ and HNCO $(10_{1,9}$--$9_{1,8})$ and
those from SO$_2$ isotopologues are only seen in MM1. We will
discuss these diversities in Section \ref{divers}.

MM1 has an integrated flux of 1.2 Jy at 1.3 mm continuum, in
agreement with the BIMA measurement of 0.97 Jy at 1.4 mm and is
completely dominated by the dust emission \citep{Gomez03}. The mm
continuum emission from MC does not appear to be very centrally
peaked and is blended with the MM2 emission. Therefore we cannot
obtain a reliable estimate of the continuum flux for MC alone. The
measured flux toward MM2 and MC amounts to 0.8 Jy and is
apparently dominated by MM2. From the continuum fluxes and
assuming thermal equilibrium between the gas and dust hence
adopting the temperatures derived from the CH$_3$CN emission
(Section \ref{divers}), we obtain a gas mass of 6--34 $M_{\odot}$
for MM1 and a total mass of 12--65 $M_{\odot}$ for MM2 and MC, for
a dust opacity $\kappa_{225{\rm GHz}}$$\sim$1.9--0.35 cm$^2$
g$^{-1}$ or an emissivity index $\beta$$\sim$1--2
\citep{Hildebrand83}.

\subsection{Molecular Outflows}
\begin{figure}
\epsscale{1.} \plotone{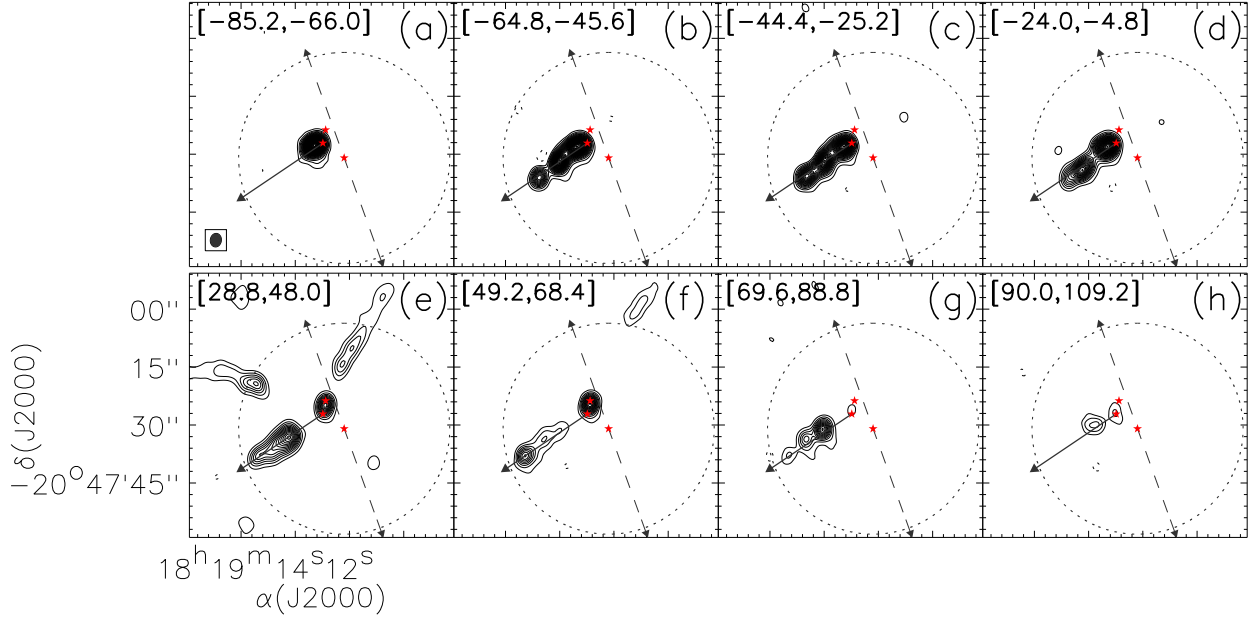} \caption{The CO (2--1) emission
integrated over every 20.4 km\,s$^{-1}$, as indicated in the upper
left of each panel, from $-85.2$ to $-4.8$ km\,s$^{-1}$ and from
28.8 to 109.2 km\,s$^{-1}$; the first and spacing contours are
0.4, 0.6, 1.2, 1.8 Jy\,beam$^{-1}$\,km\,s$^{-1}$ for panels (a)
and (h), (b) and (g), (c) and (f), (d) and (e), respectively;
hereafter the double dashed arrows delineate the orientation of a
thermal radio jet from Marti et al. (1993); the large dashed
circles mark the primary beam of the SMA. \label{co}}
\end{figure}

Figure \ref{co} shows the high velocity CO (2--1) emission
integrated over every 20.4 km\,s$^{-1}$ from $-85.2$ to $-4.8$
km\,s$^{-1}$ in the blueshifted lobe and from 28.2 to 109.2
km\,s$^{-1}$ in the redshifted lobe. The emission at lower
velocities ($|{\Delta}V|$$\lesssim$15 km\,s$^{-1}$ with respect to
the systemic velocity of 12 km\,s$^{-1}$ for MM2) suffers from
side lobes and missing flux due to inadequate $(u,v)$ coverage,
hampering the interpretation of the morphology and kinematics,
hence we focus on the high velocity CO emission in this work. The
most remarkable feature in Figure \ref{co} is a jet-like outflow
emanating from MM2 to the southeast (SE), which exhibits extremely
high velocity (EHV) emission up to $-85.2$ km\,s$^{-1}$ for the
blueshifted lobe and $103.2$ km\,s$^{-1}$ for the redshifted lobe
above a 3$\sigma$ level of $\sim$60 mJy\,beam$^{-1}$ per 1.2
km\,s$^{-1}$. This outflow shows a knotty appearance with multiple
clumps aligned along the flow axis. In addition, two more
collimated flows, the northwest and northeast flows, are seen in
Figure \ref{co}e; the former apparently originates from MC,
whereas it is difficult with the existing data to unambiguously
assign a driving source to the latter. These two flows both extend
well beyond the SMA primary beam of $\sim$54$''$ at 230 GHz, thus
are not further discussed in this work.

\begin{figure}
\epsscale{1.} \plotone{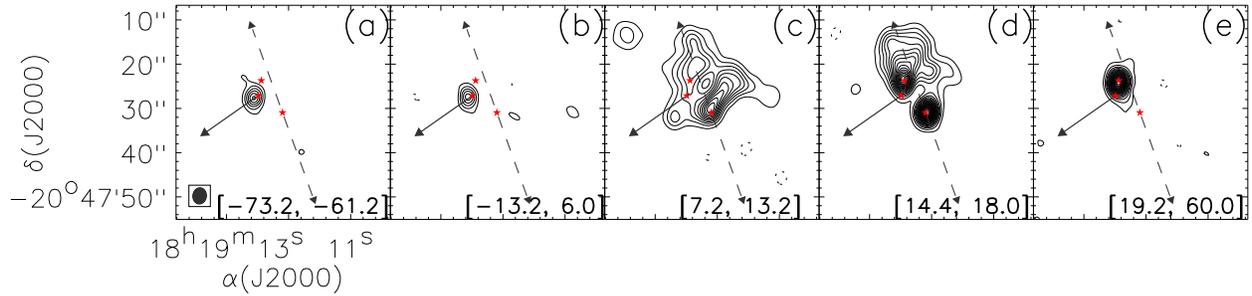} \caption{The SO ($6_5$--$5_4$)
emission, integrated over five velocity intervals as indicated in
the lower left of each panel; the first and spacing contours are
0.24, 0.3, 1.2, 0.3, 0.4 Jy\,beam$^{-1}$\,km\,s$^{-1}$ for panels
(a) to (e), respectively. \label{so}}
\end{figure}

Figure \ref{so} shows the SO ($6_5$--$5_4$) emission integrated
over five velocity intervals. The blueshifted emission arising
from the SE outflow is detected as an EHV clump in Figure
\ref{so}a and a low to high velocity clump in Figure \ref{so}b. In
Figure \ref{so}c the blueshifted low velocity emission mostly
traces a wide-angle outflow surrounding the radio jet axis. In
Figure \ref{so}d the redshifted low velocity emission reveals two
compact clumps coincident with MM1 and MC, as well as extended
structures probably attributed to outflows from MM1 and/or MC. In
Figure \ref{so}e, a redshifted clump is detected at very high
velocities; this structure can also be identified in CO in Figures
\ref{co}e and \ref{co}f, and is apparently associated with MC.

\section{Discussion}

\subsection{Physical and Chemical Diversities among the Dense
Cores} \label{divers} 
\begin{figure}
\epsscale{.8} \plotone{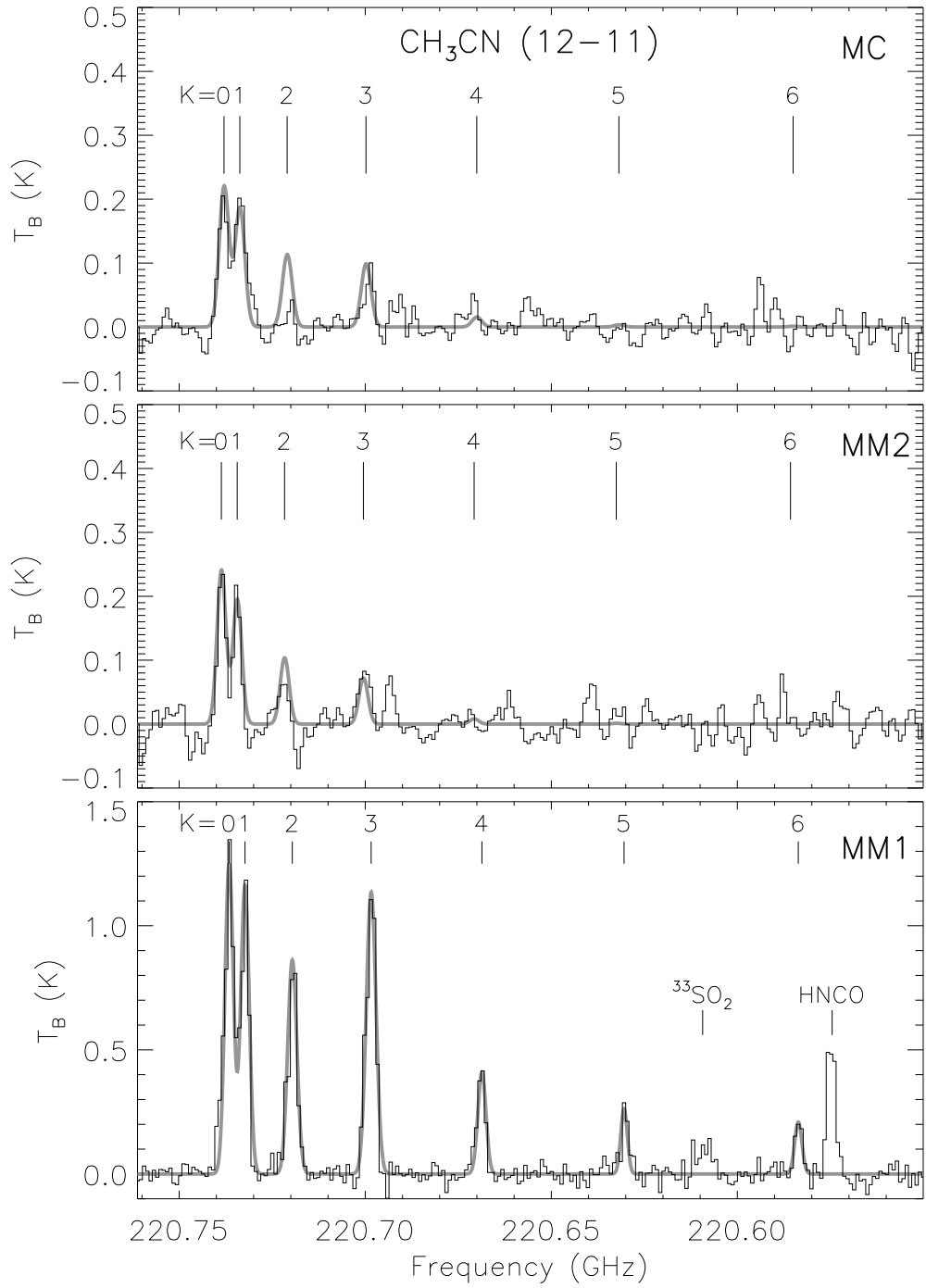} \caption{The CH$_3$CN (12--11)
spectra for MC ({\it top}), MM2 ({\it middle}), and MM1 ({\it
bottom}), taken from the positions indicated by the three stars in
Figure \ref{fig1}. The best fitting models are overlaid in thick
gray lines. \label{ch3cn}}
\end{figure}

The CH$_3$CN (12$_K$--11$_K$) emission is a
good thermometer of dense gas surrounding high-mass protostars.
Assuming that all the $K$ components are in local thermodynamical
equilibrium and tracing the same gas, we perform a simultaneous
fitting, taking into account the optical depth effect, to all the
detected components by means of grid search ${\chi}^2$
minimization. From Figure \ref{ch3cn}, the best fitting models
agree well with the observations, and yield gas temperatures of
110$_{-20}^{+30}$ K for MM1, 35$_{-15}^{+20}$ K for MM2, and
45$_{-15}^{+30}$ K for MC.

In addition to the CH$_3$OH (8$_{-1,8}$--7$_{0,7}$)$E$ line shown
in Figure \ref{fig1}b, thermal emission in CH$_3$OH
(8$_{0,8}$--7$_{1,6}$)$E$ and (3$_{-2,2}$--4$_{-1,4}$)$E$ is
detected in all the three cores but appears much brighter in MM1.
The CH$_3$OH (15$_{4,11}$--16$_{3,13}$)$E$ emission, however, is
seen in MM1 alone (Figure \ref{fig1}g). This is not surprising
since this transition has an upper energy level ($E_{up}$) of 374
K, dramatically higher than 39--96 K for the other three
transitions. As indicated by the higher temperature inferred from
the CH$_3$CN emission, only the MM1 core possesses sufficient hot
gas to populate the high-excitation CH$_3$OH.

HNCO shows a diversity similar to CH$_3$OH; the HNCO
(10$_{0,10}$--9$_{0,9}$) emission is seen in both MM1 and MC, and
probably MM2 as well, but the HNCO (10$_{1,9}$--9$_{1,8}$)
emission is detected in MM1 alone. Lines of HNCO ($J_{K_a,K_b}$)
are suggested as a good probe of the far-infrared field since its
$K_a$$>$0 ladders are populated primarily by far-infrared
radiation rather than collisions with H$_2$ \citep{Churchwell86,
Zinchenko00}. The warm/hot dust in MM1 is expected to produce
far-infrared photons responsible for the population of the $K_a$=1
HNCO, giving rise to the HNCO (10$_{1,9}$--9$_{1,8}$) emission,
however, the cooler dust in MM2 and MC would not provide
sufficient far-infrared radiation to populate appreciable HNCO at
$K_a$$>$0.

The emission from SO$_2$ and its isotopologues exhibits a
significant diversity among the three cores. Besides the SO$_2$
(11$_{5,7}$--12$_{4,8}$) emission shown in Figure \ref{fig1}i, the
lines of SO$_2$ (22$_{7,15}$--23$_{6,16}$), $^{34}$SO$_2$
(11$_{1,11}$--10$_{0,10}$), $^{34}$SO$_2$ (4$_{2,2}$--3$_{1,3}$),
$^{34}$SO$_2$ (22$_{2,20}$--22$_{1,21}$), and $^{33}$SO$_2$
(11$_{1,11}$--10$_{0,10}$), are clearly detected in MM1. However,
none of them is seen in MC or MM2. The detected transitions have
$E_{up}$ ranging from 19 to 353 K, hence the temperature variation
among the cores cannot solely account for the non-detection of any
transition in MC or MM2. In addition, the SO$_2$ chemistry does
not seem to be much dependent on luminosity or mass of the central
forming star, as bright SO$_2$ emission was detected in low-mass
star-forming regions as well \citep[e.g., IRAS
16293-2422,][]{Blake94}. It is therefore the chemical effect that
plays a central role in shaping the appearance of SO$_2$ in these
cores. If SO$_2$ forms via the reaction of
SO$+$OH$\rightarrow$SO$_2$, as adopted by \citet{Charnley97} and
\citet{Doty02}, the non-detection of SO$_2$ points to the
deficiency of OH in MC and MM2, since bright SO emission is
detected toward all the three cores. This is plausible considering
that the UV induced photodissociation of H$_2$O may be the primary
source of OH in MM1 \citep{Tappe08} and the formation of other
molecules detected in MC and MM2 does not necessarily involve the
OH radical. The rich emission lines from SO$_2$ and its
isotopologues seem to probe a more evolved stage of MM1, in
agreement with recent sulphur chemistry models \citep{Doty02,
Beuther09}.

With a gas temperature of 110 K and moderately rich molecular line
emission, MM1 seems to be at an early hot core phase
\citep{Kurtz00}. Of the two younger cores MM2 and MC, the latter
is likely more evolved with its more active molecular line
emission. One issue to be addressed toward MC is that the core
appears relatively faint in the dust continuum, which may be
caused by the deficiency of dust or by a flattened density profile
hence a large fraction of the emission being filtered out by the
interferometer. With the existing data we cannot shed more light
on this issue. However, its hot-core-like line emission is very
likely excited by a high-mass protostar. Although nearby low-mass
protostars my excite emission in complex organic molecules such as
CH$_3$OH and CH$_3$CN \citep[e.g.][]{Bottinelli04, Chandler05},
the emission may not be detectable at much larger distances.
Quantitatively, if the dust and gas reach thermal equilibrium, the
luminosity of the central source can be estimated based on the
heating of the dust \citep{Scoville76, Zhang07}. For dust
temperatures of 35--110 K at scales of $1''$ and a typical dust
emissivity index of 1.5, the luminosity of the central star ranges
from 2$\times$$10^3$ to 4$\times$$10^4$ $L_{\odot}$. Thus, the
central protostars in the three cores are likely to be massive.

\subsection{The EHV Jet-like Outflow and Molecular Bullets}
\label{bullet}

To obtain a more reasonable assessment of the mass and energetics
of the SE outflow, we calculate the gas mass down to
$|{\Delta}V|$$\sim$6 km\,s$^{-1}$; but to alleviate contamination
from ambient gas and unrelated features we only take into account
an area closely encompassing the jet-like structure. Following the
procedure of \citet{Qiu09}, we derive the outflow mass, momentum,
dynamical timescale, mass rate, and momentum rate of 0.22
$M_{\odot}$, 4.9 $M_{\odot}$ km\,s$^{-1}$, 2.2$\times$$10^3$ yr,
$10^{-4}$ $M_{\odot}$\,yr$^{-1}$, and 2.2$\times10^{-3}$
$M_{\odot}$\,km\,s$^{-1}$\,yr$^{-1}$, respectively. Considering
the missing flux and optical depth effect at low velocities, the
derived mass and energetics may represent lower limits. The
overlapping blue- and redshifted emission along the jet-like
structure suggests that the flow axis is strongly inclined to the
plane of sky. In projection the flow axis appears to wiggle over a
small angle of $\lesssim$20$^{\circ}$. If the flow axis wiggles
along the line of sight as it does in the plane of sky, the
inclination angle is likely about 10$^{\circ}$, half of the wiggle
angle. The outflow mass and momentum rates corrected for the
inclination angle would amount to 6$\times$$10^{-4}$
$M_{\odot}$\,yr$^{-1}$ and 7$\times10^{-2}$
$M_{\odot}$\,km\,s$^{-1}$\,yr$^{-1}$, respectively. The one-sided
appearance of the outflow is probably ascribed to the peculiar
environment. In projection the central source MM2 is close
($\sim$$4''$) to the HH 80--81 radio jet, which is surrounded by a
biconical cavity. The northwestern side of MM2 is likely deficient
in dense gas and the outflow is expanding into the radio jet
cavity, thus cannot create appreciable CO emission.

\begin{figure}
\epsscale{.8} \plotone{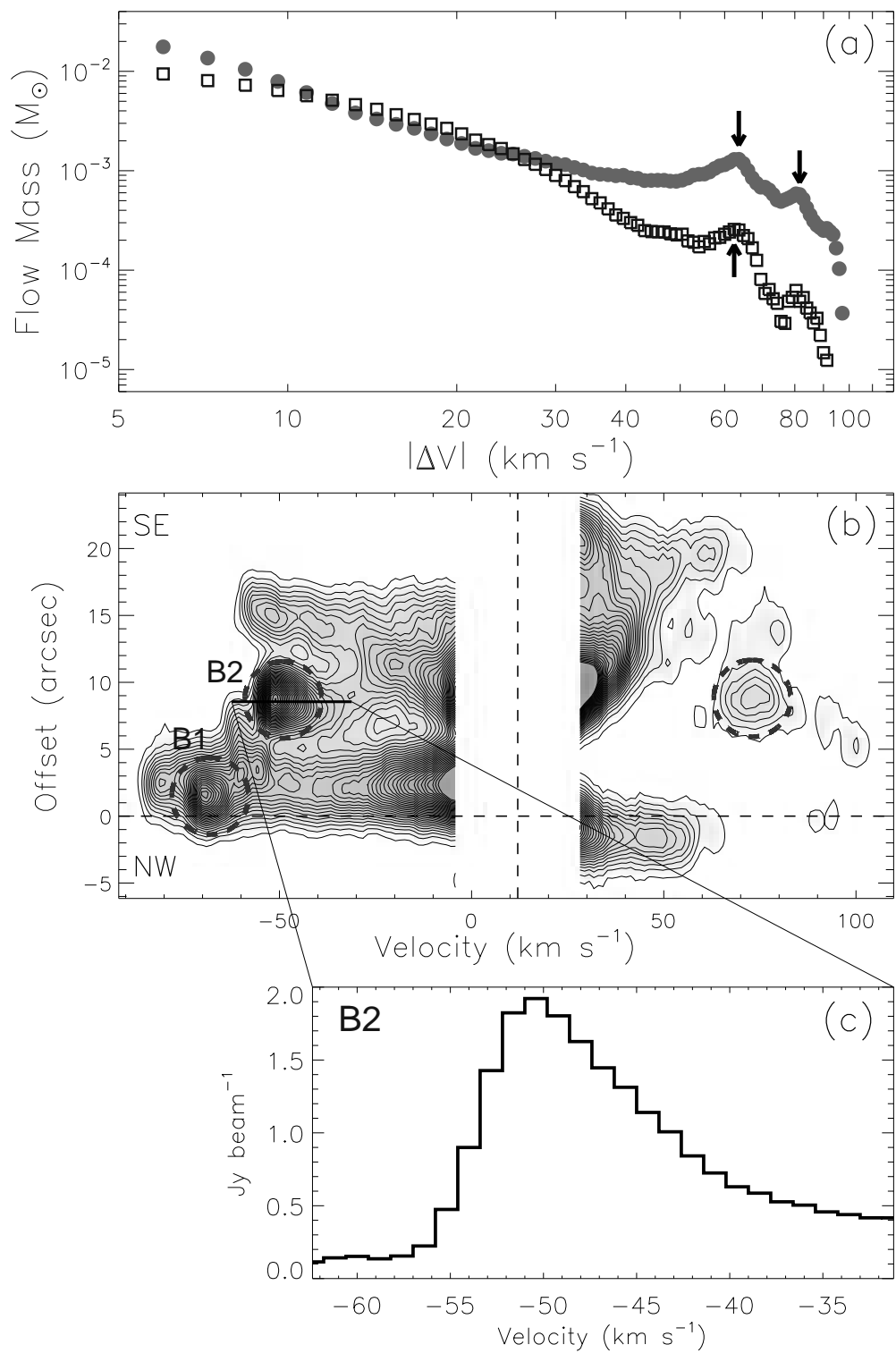} \caption{(a) Mass-velocity diagram
of the SE CO outflow, calculated in each 1.2 km\,s$^{-1}$ channel;
gray dots and dark squares denote measurements of the blue- and
redshifted lobes, respectively; three arrows mark enhancements
attributed to the molecular bullets. (b) Position-velocity diagram
along the major axis; three dashed circles highlight the molecular
bullets; the vertical dashed line denotes the systemic velocity of
the MM2 core. (c) The velocity profile of the B2 bullet.
\label{pmv}}
\end{figure}

Figures \ref{pmv}a and \ref{pmv}b show, respectively, the
mass-velocity (MV) and position--velocity (PV, cut along the major
axis) diagrams of the SE outflow. The most interesting
characteristics in the two diagrams are the distinct EHV features
that appear as abrupt bumps in the MV diagram (denoted by arrows)
and as enhanced condensations in the PV diagram (denoted by dashed
circles). This kind of compact and fast moving features are
recognized as ``molecular bullets'' in low-mass class 0
protostellar outflows \citep{Bachiller90, Hatchell99, Nisini07}.
The bullets in the SE outflow are, to our knowledge, the first
discovery of molecular bullets originating from a high-mass
star-forming core. The bright B1 and B2 bullets have sizes of
$\sim$0.04--0.05 pc and masses of 10$^{-3}$ $M_{\odot}$. However,
recent multi-transition SiO and CO observations yield hot gas
($>$100 K) in low-mass molecular bullets \citep{Hatchell99,
Nisini07}; if this is the case for the B1 and B2 bullets, their
masses would reach $\gtrsim$10$^{-2}$ $M_{\odot}$. It is also
worth noting that the B1 bullet is clearly detected in SO
($6_5$--$5_4$), suggesting shock activity in this bullet (Figure
\ref{so}a).

From its velocity profile shown in Figure \ref{pmv}c, the B2
bullet has the bulk material moving at highest velocities, with a
gradual decrease toward the systemic velocity. The B1 bullet
exhibits a similar but less prominent profile. The velocity
structure of the two bullets is in direct contrast to what is
expected if the bullets were created in situ from entrained or
swept-up ambient medium, in which case the profile would have a
steep decrease toward the systemic velocity
\citep[e.g.,][]{Zhang95, Qiu07}. It is more likely that the
majority of the bullet material is ejected from the close vicinity
of the central protostar. In this scenario the presence of
multiple molecular bullets directly manifests an episodic nature
of the outflow.

\citet{Marti99} resolved H$_2$O maser spots toward MM2 into two
groups, with one group aligned in a linear distribution and the
other located to the NE. Although the H$_2$O maser spots are seen
very close to MM2 ($\sim$500 AU in projection), we find that in
both velocity and orientation the NE and linear maser groups agree
well with the B1 and B2 bullets, respectively. This further
supports the scenario that the bullets are ejected from the close
vicinity of the central protostar. A sketchy and tempting
interpretation is that the episodic ejections from the central
protostar or star-disk system collides with the existing flow,
re-accelerating the flow material to form the fast moving bullets;
the maser emission, which is pumped under critical conditions,
signifies the collision sites. Future proper motion measurements
of the H$_2$O maser spots and multi-transition observations of the
bullets may place more stringent constraints on the excitation
mechanisms of the two extreme phenomena.

High velocity, jet-like molecular outflows are often seen in
low-mass class 0 protostars \citep[e.g.,][]{Guilloteau92,
Bachiller95, Gueth99, Lee07}; in theoretical models they are
physically linked to the earliest-phase disk accretion within a
dense envelope \citep{Shang06, Fendt09}. However, jet-like
outflows with high velocity emission are still rarely identified
toward high-mass star-forming regions \citep[e.g.,][]{Qiu07}.
Based on single-dish CO observations, \citet{Rodriguez-Franco99}
report high velocity bullets and jets toward the well-known
Orion-KL outflow, but recent interferometric observations show a
fan-shaped or explosive morphology of the outflow
\citep{Beuther08, Zapata09}. The SE outflow has a well-defined
jet-like morphology, reaching an order of 100 km\,s$^{-1}$ in both
blue- and redshifted lobes, thus apparently favors a disk-mediated
accretion onto the central protostar. The EHV emission may trace
part of the outer layer of an underlying wind. Moreover, the
outflow contains multiple molecular bullets, indicating an
episodic mass-loss process and further pointing to an episodic
nature of the central accretion process.

\section{Summary}
We report the discovery of jet-like CO outflows in the HH 80--81
star-forming region. In particular, the outflow arising from core
MM2 exhibits blue- and redshifted line wings close to 100
km\,s$^{-1}$ from the cloud velocity, and molecular bullets, which
points to an episodic, disk-mediated accretion for massive star
formation. Three dense cores identified in continuum and molecular
line emission demonstrate physical and chemical diversities which
suggest an evolutionary sequence that MM2 is younger than MC,
which is in turn younger than the hot molecular core MM1.

\acknowledgments We thank Ciriaco Goddi for discussions on water
and methanol maser emission in high-mass star-forming regions.

\begin{thebibliography}{}
\bibitem[Aspin et al.(1994)]{Aspin94}Aspin, C. et al. 1994, \aap, 292, L9
\bibitem[Bachiller et al.(1990)]{Bachiller90}Bachiller, R.,
Cernicharo, J., Mart\'{\i}n-Pintado, J., Tafalla, M., \& Lazareff,
B. 1990, \aap, 231, 174
\bibitem[Bachiller et al.(1995)]{Bachiller95}Bachiller, R., Guilloteau,
S., Dutrey, A., Planesas, P., \& Martin-Pintado, J. 1995, \aap,
299, 857
\bibitem[Beuther \& Nissen(2008)]{Beuther08}Beuther, H. \& Nissen,
H. D. 2008, \apj, 679, 121L
\bibitem[Beuther et al.(2002)]{Beuther02}Beuther, H., Schilke,
P., Gueth, F., McCaughrean M., Andersen, M., Sridharan, T. K., \&
Menten, K. M. 2002, \aap, 387, 931
\bibitem[Beuther et al.(2009)]{Beuther09}Beuther, H., Zhang, Q.,
Bergin, E. A., \& Sridharan, T. K. 2009, \aj, 137, 406
\bibitem[Blake et al.(1994)]{Blake94}Blake, G. A., van Dishoek, E.
F., Jansen, D. J., Groesbeck, T. D., \& Mundy, L. G. 1994, \apj,
428, 680
\bibitem[Bottinelli et al.(2004)]{Bottinelli04}Bottinelli, S. et al.
2004, \apj,  615, 354
\bibitem[Cesaroni et al.(2005)]{Cesaroni05}Cesaroni, R., Neri, R.,
Olmi, L., Testi, L., Walmsley, C. M., \& Hofner, P. 2005, \aap,
434, 1039
\bibitem[Chandler et al.(2005)]{Chandler05}Chandler, C. J.,
Brogan, C. L., Shirley, Y. L., \& Loinard, L. 2005, \apj, 632, 371
\bibitem[Charnley (1997)]{Charnley97}Charnley, S. B. 1997, \apj,
481, 396
\bibitem[Churchwell et al.(1986)]{Churchwell86}Churchwell, E.,
Wood, D., Myers, P. C., \& Myers, R. V. 1986, \apj, 305, 405
\bibitem[Doty et al.(2002)]{Doty02}Doty, S. D., van Dishoek, E.
F., van der Tak, F. F. S., \& Boonman, A. M. S. 2002, \aap, 389,
446
\bibitem[Fendt(2009)]{Fendt09}Fendt, C. 2009, \apj, 692, 346
\bibitem[G\'{o}mez et al.(2003)]{Gomez03}G\'{o}mez, Y.,
Rodr\'{\i}guez, L. F., Girart, J. M., Garay, G., \& Mart\'{\i}, J.
2003, \apj, 597, 414
\bibitem[G\'{o}mez et al.(1995)]{Gomez95}G\'{o}mez, Y.,
Rodr\'{\i}guez, L. F., \& Mart\'{\i}, J. 1995, \apj, 453, 268
\bibitem[Gueth \& Guilloteau(1999)]{Gueth99}Gueth, F. \& Guilloteau,
S. 1999, \aap, 343, 571
\bibitem[Guilloteau et al.(1992)]{Guilloteau92}Guilloteau, S.,
Bachiller, R., Fuente, A., \& Lucas, R. 1992, \aap, 265, L49
\bibitem[Hatchell et al.(1999)]{Hatchell99}Hatchell, J., Fuller,
G. A., \& Ladd, E. F. 1999, \aap, 346, 278
\bibitem[Hildebrand (1983)]{Hildebrand83}Hildebrand, R. H. 1983,
\qjras, 24, 267
\bibitem[Keto \& Wood(2006)]{Keto06}Keto, E. \& Wood, K. 2006,
\apj, 637, 850
\bibitem[Kurtz et al.(2000)]{Kurtz00}Kurtz, S., Cesaroni, R.,
Churchwell, E., Hofner, P., \& Walmsley, C. M. 2000, in Protostars
\& Planets IV, ed. Mannings, V., Boss, A. P., \& Russell, S. S.
(Tucson, AZ: Univ. Arizona Press), 299
\bibitem[Kurtz et al.(2004)]{Kurtz04}Kurtz, S., Hofner, P., \&
\'{A}lvarez, C. V. 2004, \apjs, 155, 149
\bibitem[Lee et al.(2007)]{Lee07}Lee, C.-F., Ho, P. T. P., Palau,
A., Hirano, N., Bourke, T., L., Shang, H., \& Zhang, Q. 2007,
\apj, 670, 1188
\bibitem[Mart\'{\i} et al.(1993)]{Marti93}Mart\'{\i}, J.,
Rodgr\'{\i}guez, L. F., \& Reipurth, B., 1993, \apj, 416, 208
\bibitem[Mart\'{\i} et al.(1995)]{Marti95}Mart\'{\i}, J.,
Rodgr\'{\i}guez, L. F., \& Reipurth, B., 1995, \apj, 449, 184
\bibitem[Mart\'{\i} et al.(1998)]{Marti98}Mart\'{\i}, J.,
Rodgr\'{\i}guez, L. F., \& Reipurth, B., 1998, \apj, 502, 337
\bibitem[Mart\'{\i} et al.(1999)]{Marti99}Mart\'{\i}, J.,
Rodgr\'{\i}guez, L. F., \& Torrelles, J. M. 1999, \aap, 345, L5
\bibitem[Nisini et al.(2007)]{Nisini07}Nisini, B., Codella, C.,
Giannini, T., Santiago Garcia, J., Richer, J. S., Bachiller, R.,
\& Tafalla, M. 2007, \aap, 462, 163
\bibitem[Qiu et al.(2008)]{Qiu08}Qiu, K., et al. 2008, \apj, 685, 1005
\bibitem[Qiu et al.(2007)]{Qiu07}Qiu, K., Zhang, Q., Beuther, H.,
\& Yang, J. 2007, \apj, 654, 361
\bibitem[Qiu et al.(2009)]{Qiu09}Qiu, K., Zhang, Q., Wu, J., \&
Chen, H. 2009, \apj, 696, 66
\bibitem[Rathborne et al.(2008)]{Rathborne08}Rathborne, J. M.,
Jackson, J. M., Zhang, Q., \& Simon, R. 2008, ApJ, 689, 1141
\bibitem[Rodr\'{\i}guez et al.(1980)]{Rodriguez80}Rodr\'{\i}guez, L. F.,
Moran, J. M., Ho, P. T. P., \& Gottlieb, E. W. 1980, \apj, 235,
845
\bibitem[Rodr\'{\i}guez-Franco et
al.(1999)]{Rodriguez-Franco99}Rodr\'{\i}guez-Franco, A.,
Mart\'{\i}n-Pintado, J., \& Wilson, T. L. 1999, \aap, 344, L57
\bibitem[Scoville \& Kwan(1976)]{Scoville76}Scoville, N. Z. \&
Kwan, J. 1976, \apj, 206, 718
\bibitem[Shang et al.(2006)]{Shang06}Shang, H., Allen, A. Li,
Z.-Y., Liu, C.-F., Chou, M.-Y., Anderson, J. 2006, \apj, 649, 845
\bibitem[Shepherd et al.(1998)]{Shepherd98}Shepherd, D. S.,
Watson, A. M., Sargent, A. I., \& Churchwell, E. 1998, \apj, 507,
861
\bibitem[Stecklum et al.(1997)]{Stecklum97}Stecklum, B., Feldt,
M., Richichi, A., Calamai, G., \& Lagage, P. O. 1997, \apj, 479,
339
\bibitem[Slysh et al.(2002)]{Slysh02}Slysh, V. I., Kalenskii, S.
V., \& ValTts, I. E. 2002, Astronomy Reports, 46, 49
\bibitem[Tappe et al.(2008)]{Tappe08}Tappe, A., Lada, C. J.,
Black, J. H., \& Muench, A. A. 2008, \apj, 680, L117
\bibitem[Zapata et al.(2009)]{Zapata09}Zapata, L., Rodr\'{\i}guez,
L. F., Ho, P., Menten, K., \& Schmid-Burgk, J. 2009, poster
presented at the Millimeter and Submillimeter Astronomy at High
Angular Resolution meeting, Taipei, Taiwan
\bibitem[Zhang et al.(1995)]{Zhang95}Zhang, Q., Ho., P. T. P.,
Wright, M. C. H., \& Wilner, D. J. 1995, \apj, 451, L71
\bibitem[Zhang et al.(2007)]{Zhang07}Zhang, Q., Hunter, T. R.,
Beuther, H., Sridharan, T. K., Liu, S.-Y., Su, Y.-N., Chen, H.-R.,
\& Chen, Y. 2007, \apj, 658, 1152
\bibitem[Zhang et al.(1998)]{Zhang98}Zhang, Q., Hunter, T. R., \&
Sridharan, T. K. 1998, \apj, 505, L151
\bibitem[Zhang et al.(2009)]{Zhang09}Zhang, Q., Wang, Y., Pillai, T.,
Rathborne, J. 2009, \apj, 696, 268
\bibitem[Zinchenko et al.(2000)]{Zinchenko00}Zinchenko, I.,
Henkel, C., \& Mao, R. Q. 2000, \aap, 361, 1079
\end{thebibliography}
\end{document}